\documentclass[pra,aps,showpacs,preprint]{revtex4}
\usepackage{dcolumn}

\begin{document}

\input epsf

\def\WIMPZILLA{{\sc wimpzilla}}
\def\WIMPZILLAS{{\sc wimpzillas}}
\def\WIMP{{\sc wimp}}
\def\WIMPS{{\sc wimps}}
\def\mpl{M_{Pl}}
\def\simlt{\stackrel{<}{{}_\sim}}
\def\simgt{\stackrel{>}{{}_\sim}}

\title{On the gravitational production of superheavy dark matter}  

\author{Daniel J.\ H.\ Chung}\thanks{Electronic mail: djchung@umich.edu}
	\affiliation{Michigan Center for Theoretical Physics,
     	University of Michigan, Ann Arbor, Michigan \ 48109}

\author{Patrick\ Crotty}\thanks{Electronic mail: prcrotty@oddjob.uchicago.edu}
	\affiliation{Department of Physics, Enrico Fermi Institute, \\
     	The University of Chicago, Chicago, Illinois \ 60637-1433}

\author{Edward W.\ Kolb}\thanks{Electronic mail: rocky@rigoletto.fnal.gov}
	\affiliation{NASA/Fermilab Astrophysics Center, Fermi
     	National Accelerator Laboratory, Batavia, Illinois \  60510-0500,\\
        and Department of Astronomy and Astrophysics, Enrico Fermi Institute,\\
     	The University of Chicago, Chicago, Illinois \ 60637-1433}

\author{Antonio Riotto}\thanks{Electronic mail:
antonio.riotto@pd.infn.it}
	\affiliation{INFN, Sezione di Padova, Via Marzolo 8, Padova 
	I-35131, Italy}

\date{March, 2001}
\begin{abstract}
The dark matter in the universe can be in the form of a superheavy matter
species (\WIMPZILLA).  Several mechanisms have been proposed for the production
of \WIMPZILLA\ particles during or immediately following the inflationary
epoch.  Perhaps the most attractive mechanism is through gravitational particle
production, where particles are produced simply as a result of the expansion of
the universe.  In this paper we present a detailed numerical calculation of
\WIMPZILLA\ gravitational production in hybrid-inflation models and
natural-inflation models. Generalizing these findings, we also explore the
dependence of the gravitational production mechanism on various models of
inflation. We show that superheavy dark matter production seems to be robust,
with $\Omega_Xh^2 \sim (M_X/10^{11}\textrm{GeV})^2 (T_{RH}/10^9\textrm{GeV})$,
so long as $M_X<H_I$, where $M_X$ is the \WIMPZILLA\ mass, $T_{RH}$ is the
reheat temperature, and $H_I$ is the expansion rate of the universe during
inflation.
\end{abstract}

\pacs{98.80.Cq, 95.35.+d, 4.62.+v; FERMILAB-Pub-01/047-A; MCTP-01-16; 
hep-ph/0104100}

\maketitle

\section{Introduction}

The case for dark, nonbaryonic matter in the universe is today stronger than
ever \cite{turner}.  The observed large-scale structure suggests that dark
matter (DM) accounts for at least 30\% of the critical mass density of the
universe $\rho_C=3 H_0^2 \mpl^2/8\pi = 1.88\times 10^{-29}$ g cm$^{-3}$, where
$H_0\equiv 100 h$ km\,sec$^{-1}$ Mpc$^{-1}$ is the present Hubble constant and
$\mpl$ is the Planck mass.

Despite this compelling evidence, the nature of the DM is still unknown. Some
fundamental physics beyond the Standard Model (SM) is certainly required to
account for the cold and slowly moving particles, $X$, composing the the bulk
of the nonbaryonic dark matter.  

The most familiar assumption is that dark matter is a thermal relic, {\it
i.e.,} it was initially in chemical equilibrium in the early universe.  A
particle species, $X$, tracks its equilibrium abundance as long as reactions
which keep the species in chemical equilibrium can proceed on a timescale more
rapid than the expansion rate of the universe, $H$.  When the reaction rate
becomes smaller than the expansion rate, the particle species can no longer
track its equilibrium value.  When this occurs the particle species is said to
be ``frozen out.''  The more strongly interacting the particle, the longer it
stays in local thermal equilibrium and the smaller its eventual freeze-out
abundance.  Conversely, the more weakly interacting the particle, the larger
its present abundance.  If freeze out occurs when the particles $X$ are
nonrelativistic, the freeze-out value of the particle number per comoving
volume $Y$ is related to the mass of the particle and its annihilation cross
section (here characterized by $\sigma_0$) by \cite{book} $ Y \propto (1/M_X
\mpl \sigma_0)$ where $M_X$ is the mass of the particle $X$.  Since the
contribution to $\Omega_X=\rho_X/\rho_C$ is proportional to $M_Xn_X$, which in
turn is proportional to $M_XY$, the present contribution to $\Omega_X$ from a
thermal relic roughly is {\em independent} of its mass and depends only upon
the annihilation cross section.  The cross section that results in
$\Omega_Xh^2\sim 1$ is of order $10^{-37}$cm$^2$, which is of the order the
weak scale.  Many theories beyond the SM, {\it e.g.} supersymmetric theories,
have stable particles with weak-scale annihilation cross sections, and provide
candidate weakly interacting massive particles (\WIMPS).

The simple assumption that dark matter is a thermal relic limits the maximum
mass of the DM.  The largest possible annihilation cross section is roughly
$M_X^{-2}$.  This implies that very massive \WIMPS\ would have such a small
annihilation cross section that their present abundance would be too large.
Thus, one expects a maximum mass for a thermal \WIMP, which turns out to be a
few hundred TeV \cite{griestkam}.

One should note that the computation of the final abundance of the thermal
relics assumes that the largest temperature of the universe was larger
than the relic
mass $M_X$. The thermal history of the universe before the epoch of
nucleosynthesis is unknown, and the maximum temperature in the
radiation-dominated phase, dubbed the reheating temperature ($T_{RH}$), might
have been smaller than the mass of the \WIMP. In such a case, the dependence of
the present abundance on the mass and the annihilation cross section differs
from familiar results because of the new parameter $T_{RH}$ \cite{gkr}. This
drastically changes the cosmologically allowed parameter space of
supersymmetric models and re-establishes SM neutrinos as possible dark matter
candidates \cite{gkrst}.

While a thermal origin for \WIMPS\ is the most common assumption, it is not the
simplest possibility.  It has been recently pointed out that DM particles might
have never experienced local chemical equilibrium during the evolution of the
universe, and that their mass may be in the range $10^{12}$ to $10^{19}$ GeV,
much larger than the mass of thermal \WIMPS\ \cite{ckr1,ckr2,ckr3,ckr4}. Since
these \WIMPS\ would be much more massive than thermal \WIMPS, such superheavy
DM particles have been called \WIMPZILLAS\ \cite{ckr4}.

Since \WIMPZILLAS\ are extremely massive, the challenge lies in creating very
few of them. Several \WIMPZILLA\ scenarios have been developed involving
production during different stages of the evolution of the universe. 

\WIMPZILLAS\ may be created during bubble collisions if inflation is completed
through a first-order phase transition \cite{barrow,mr}; at the preheating
stage after the end of inflation with masses easily up to the Grand
Unified scale of
$10^{15}$GeV  \cite{Kolb:1998he} or even  up to the Planck scale
 \cite{Giudice:1999fb}; or during the reheating stage after
inflation \cite{ckr3} with masses which may be as large as $2\times 10^3$ times
the reheat temperature.

\WIMPZILLAS\ may also be generated in the transition between an inflationary
and a matter-dominated (or radiation-dominated) universe due to the
``nonadiabatic'' expansion of the background spacetime acting on the vacuum
quantum fluctuations. This mechanism was studied in details in Refs.\
\cite{ckr1,kuzmin} in the case of chaotic inflation. The distinguishing feature
of this mechanism is the capability of generating particles with mass of the
order of the inflaton mass (usually much larger than the reheating temperature)
even when the particles only interact extremely weakly (or not at all) with
other particles, and do not couple to the inflaton.

While the results depend weakly on details such as whether
the \WIMPZILLA\ is a fermion or a boson, or whether it is conformally or
minimally coupled to gravity, for the most part $\Omega_X \sim 1$ when the mass
of the \WIMPZILLA\ is approximately the order of the inflaton mass.  Since
hybrid inflation models have (at least) two mass scales and more coupling
constants than chaotic inflation models, it is worthwhile to study \WIMPZILLA\
production in hybrid models \cite{hybrid}.

In this paper we study the gravitational production of \WIMPZILLAS\ after the
completion of a stage of hybrid inflation.  The hybrid scenario involves two
scalar fields, the inflaton field $\phi$, and the symmetry-breaking field
$\sigma$. Models are parameterized by different mass scales and couplings for
the two fields. During inflation the inflaton field $\phi$ rolls down along a
flat potential while the field $\sigma$ is stuck at the origin, providing the
vacuum energy density driving inflation. However, when $\phi$ becomes smaller
than a critical value, $\phi_c$, both fields roll down very quickly towards
their present minima, completing the inflationary phase.  It is exactly during
this phase the gravitational generation of
\WIMPZILLAS\ may occur.

If the \WIMPZILLAS\ are produced at the end of inflation, the fraction of the
total energy density of the universe in \WIMPZILLAS\ today is given by
\begin{equation} 
\Omega_X h^2 \approx \Omega_R h^2\:
\left(\frac{T_{RH}}{T_0}\right)\: 
\frac{8 \pi}{3} \left(\frac{M_X}{\mpl}\right)\:
\frac{n_X(t_{e})}{\mpl H_I^2},
\label{eq:omegachi}
\end{equation}
where $H_I$ is the expansion rate of the universe at the end of inflation.
Here, $\Omega_R h^2 \approx 4.31 \times 10^{-5}$ is the fraction of critical
energy density in radiation today, $T_0$ is the present temperature of
radiation, and $n_X(t_{e})$ is the density of $X$ particles at the time when
they were produced. The present abundance of the nonthermal \WIMPZILLAS\ is, as
expected, independent of the cross section \cite{ckr1,ckr2}, and one can easily
verify that if there is some way to create \WIMPZILLAS\ in the correct
abundance to give $\Omega_X\sim 1$, nonequilibrium during the evolution of the
universe is automatic.

The paper is organized as follows. In Section II we present some details of the
simplest hybrid inflation model and discuss the allowed range of the various
parameters. In Section III we present our analytical results for
\WIMPZILLA\ production, making use of some general results presented in the
appendix. Section IV contains our numerical results.  Finally, in Section V we
present our conclusions.

\section{The hybrid inflation model}

For our computation of \WIMPZILLA\ production, we take the simplest hybrid
inflation potential as suggested by Linde \cite{hybrid}
\footnote{For other hybrid inflation models, including those motivated
by supersymmetry, see \cite{lr}.}
\begin{equation}
\label{Linde_potential}
V(\phi,\sigma) = \frac{1}{4\lambda}\left(m_\sigma^2-\lambda \sigma^2\right)^2
+ \frac{1}{2} m_\phi^2 \phi^2 + \frac{1}{2} g^2 \phi^2 \sigma^2\ .
\end{equation}
This potential has a valley of minima at $\sigma = 0$ for $\phi > \phi_c
\equiv m_\sigma/g$.  Most of inflation occurs while $\phi$
is slowly rolling down from its initial value to $\phi_c$.

During inflation $\sigma$ has a minimum at $\sigma =0$ and its kinetic energy
is quickly damped by the Hubble expansion. Hence, classically in this naive
picture, $\sigma$ remains at 0 for a long time before it falls due to some
infinitesimal residual displacement of $\sigma$ and/or $\dot{\sigma}$ about 0
\footnote{Because $\sigma=0$ is an unstable point, the time length before
falling is proportional to the logarithm of the inverse residual
displacement.}.  However, this picture is valid, strictly speaking, only when
one neglects quantum fluctuations. Physically, what will occur is that the
quantum fluctuations will grow and the long wavelength modes will condense such
that different regions of spacetime will behave as if they had a classical
scalar field value of $\sigma = \pm m_\sigma/\sqrt{\lambda}$ with domain walls
between the plus and minus regions. (In the case that the scalar field is
complex, a cosmic string will form instead of a domain wall.) This phenomenon
is sometime called spinodal decomposition. 

A relevant observation for gravitational particle production is that the
effective stress caused by the field gradients will increase the pressure of
the universe such that the Hubble expansion will slow faster. One way to see
this is to note that the energy conservation equation
\begin{equation} 
d\left(\rho a^{3}\right)=-Pd(a^{3})
\end{equation} 
tells us that
\begin{equation} 
\rho =\rho_{i}\left[
\frac{a_{i}}{a}\right] ^{3}-\frac{1}{a^{3}}\int _{a_{i}}^{a}Pd\left[
\frac{a}{a_{i}}\right] ^{3},
\end{equation} 
which implies that a positive increase in the pressure will lead to a faster
decrease in the energy density, causing a faster decrease in $H$. Of course,
even if the universe contains inhomogeneities due to these field gradients, one
can average over the fluctuation to account for an effective energy density and
pressure.

One way of accounting for quantum fluctuations has been presented by
Ref.~\cite{usefulspinodal}.  There, the canonical formalism is used to quantize
the fluctuations about a time dependent zero mode $\bar{\sigma }(t)$: $\sigma
=\bar{\sigma }(t)+\delta \sigma (x,t)$.  They argue that the long wavelength
modes of $\delta \sigma (x,t)$ condense such as to form an effectively
homogeneous scalar field $\delta \bar{\sigma}(t)$, whose energy contribution to
the stress energy tensor can dominate over the stress energy of the background
mode $\bar{\sigma}(t)$ such that the expansion rate $\dot{a}/a$ is damped more
quickly than one would naively expect from accounting for only
$\bar{\sigma}(t)$. This effectively homogeneous scalar field $\delta
\bar{\sigma}(t)$ has an initial condition that is fixed by $\langle \delta
\sigma^2(x,t)\rangle$ in the background of $\bar{\sigma}(t)$. It is
\begin{equation}
\langle \delta\bar{\sigma}^2(t_0)\rangle^{1/2}\approx \frac{H_I}{2\pi },
\end{equation}
where the exact numerical factor depends on the boundary condition of
the quantum fluctuations (which cannot be zero due to canonical commutation
relations), and $H_I$ is the Hubble expansion rate during inflation.

We will implement this result and simulate the condensation $\delta 
\bar{\sigma}$ and its fall by letting $\sigma$ have a nonzero initial 
condition at the end of inflation with a value of order $H_I/2\pi$ and letting
it fall, instead of having the condensation component fall. To achieve this, we
add a perturbation potential
\begin{equation}
\label{eq:perturb}
V_P(\phi ,\sigma )=BH_I^3
\left(\sigma -\frac{m_{\sigma}}{\sqrt{\lambda}} \right)\,
\exp\left[-C\left(\phi -\phi _{c}\right)^{2}\right].
\end{equation}
Then, by adjusting $B$ and $C$ we can simulate the condensate $\delta
\bar{\sigma}$ by making $\sigma(t)$ roll to the new minimum instead. We
shall, however, not take into account the potential for $\delta
\bar{\sigma}(t)$ as is done in Ref.~\cite{usefulspinodal}. In detail, if the
potential for $\sigma$ is as given in Eq.\ (\ref{Linde_potential}), the
potential in which $\delta \bar{\sigma}$ falls would be
\begin{equation} 
V(\phi ,\sigma)=\frac{1}{4\lambda }\left(m_{\sigma }^{2}-\lambda 
\bar{\sigma}^{2}\right)^{2}+\frac{1}{2}m_{\phi }^{2}\phi ^{2}
+\frac{1}{2}g^{2}\phi^{2}\bar{\sigma }^{2}
+\frac{1}{2}\left(-m_{\sigma }^{2}+g^{2}\phi^{2}
+3\lambda \bar{\sigma }^{2}\right)\delta \bar{\sigma
}^{2}+\frac{3\lambda }{4}\delta \bar{\sigma }^{4},
\label{eq:fluctuationpotential}
\end{equation}
where \(\bar{\sigma }=0$ in our case. Comparing this expression with the
tree-level effective potential, one finds that the potential for $\sigma$ with
a slight displacement from $\sigma =0$ achieves the same dynamics as $\delta
\bar{\sigma}(t)$ if $\lambda$ is replaced with $3\lambda$.  Hence, if we only
consider the case where $\sigma=0$ forever without the quantum fluctuations,
our simulated treatment of spinodal decomposition will coincide with that of
Ref.~\cite{usefulspinodal} with just the reinterpretation of
$\lambda\rightarrow 3\lambda$.  On the other hand, in reality, since $\sigma$
will never precisely be at zero forever even in the nonrealistic absence of
quantum fluctuations, a better simulated treatment of the spinodal
decomposition requires further modifications of the potential along the lines
of Eq.\ (\ref{eq:fluctuationpotential}) with $\bar{\sigma} \neq 0$. Since we
are primarily concerned with order of magnitude accuracy, and since this
approximation neglects classical wave scattering effects taken into account in
Ref.~\cite{lattice}, we will not account for this effective change in the
potential for $\sigma$.

Let us be more precise about the order of magnitude of $B$ and $C$.  To
displace effectively $\sigma$ by $H_I/2\pi$ at the end of inflation, we must
have
\begin{equation}
B\approx \frac{10^{7}g^{2}}{\lambda}\left[ \frac{m_{\sigma}}{\mpl}\right]^2
\frac{1}{\ln [1+g/\sqrt{C}m_{\sigma}]^{2}},
\end{equation}
where we have used the COBE determination of curvature perturbations, 
giving rise to the relationship \footnote{To
obtain this estimate for $B$, one integrates the equation of motion for
$\sigma$ due to the force from the potential Eq.\ (\ref{eq:perturb}) starting
from the time when $C(\phi-\phi_c)^2$ becomes order 1 and $\phi$ obeying the
equation of motion for a slowly rolling scalar field.}
\begin{equation}
m^2_\phi \approx \frac{g}{\lambda^{3/2}}\ \frac{m_\sigma^5}
{3.5\times 10^{-5}\mpl^3}.
\end{equation}  
Note that the precise value of $B$ and $C$ will not be important to the
determination of the Bogoliubov coefficient as long as the perturbation
potential causes $\sigma$ to fall. We have checked this numerically as
shown below in the case where we have set $m_{\sigma }=10^{-3}\mpl$.

We would like to emphasize that while our treatment of spinodal decomposition
is adequate for the purposes at hand, it is far from complete. Since
Ref.~\cite{lattice} argues that generically hybrid inflation ends after one
oscillation, we cannot realistically probe the parameter space in our model
where more than one oscillation of the scalar fields is important if we neglect
the important pressure-related effects due to condensation and classical-wave
scattering.  Even for the one oscillation approximation, the effect of
neglecting the pressure due to condensation and classical-wave scattering
underestimates particle production due to the fact that the pressure effects
increase the nonadiabticity of the expansion of spacetime. Hence, this issue
certainly deserves more investigation.  We note that other related references
include
Refs.~\cite{otherspinodal1,otherspinodal2,otherspinodal3,otherspinodal4} and
references therein.

The parameters in the potential in Eq.\ (\ref{Linde_potential}) are constrained
by several considerations.  Constraints on the amplitude and the tilt of the
curvature perturbation spectrum generated during inflation impose the following
constraints on $\lambda$ and $g$ \cite{lr}:
\begin{equation}
\label{mphi_inequality}
\frac{g}{\lambda^{3/2}} \frac{m_\sigma^5}{m_\phi^2 \mpl^3} \approx   3.5 
\times 10^{-5},
\end{equation}
and
\begin{equation}
\label{lain}
\frac{\lambda m_\phi^2 \mpl^2}{\pi m_\sigma^4} \simlt  0.25.
\end{equation}
The requirement that the cosmological constant term dominates during the
inflationary regime above $\phi_c$ imposes a third constraint,
\begin{equation}
\label{other_mphi_inequality}
m_\phi^2 \ll \frac{g^2 m_\sigma^2}{\lambda}.
\end{equation}
Note that the tilt of the curvature perturbation spectrum yields a
constraint similar to the condition that the $\phi$ field evolution is
slow roll; {\it i.e.,}
\begin{equation}
\frac{m_\sigma^2}{m_\phi \mpl} \gg \sqrt{\frac{3\lambda}{2\pi}} .
\end{equation}
Also, note that the condition that the cosmological constant term
dominates during the inflationary regime with $\phi>\phi_c$ also
implies the ``waterfall'' condition (the condition that the scalar
fields after $\phi$ reaches $\phi_c$ roll to the new minima quickly
compared to the expansion rate). 

With $m_\sigma$ fixed, these constraints collectively determine a region of
$(g,\lambda)$ parameter space, outside of which is forbidden by the
perturbation amplitude and tilt considerations.  Yet there is one other
constraint that we have not discussed.  As we have reviewed previously, our
model does not describe the evolution of the expansion rate of the universe
accurately beyond one oscillation of the scalar fields after the end of
inflation.  As we will see in the next section, our relic density will depend
upon an accurate modeling of the background equation for at least one Hubble
time at the end of inflation.  Hence, our model is valid only in the regime in
which no more than one oscillation takes place during one Hubble time.  Let us
see how this constrains our parameter space.

The time scale for the scalar field oscillation is set by the mass matrix (in
the $(\sigma ,\phi )$ basis)
\begin{equation} 
m^2(t)=     \frac{1}{2}     \left( \begin{array}{cc} 
-m_\sigma^2 + g^2\phi^2 + 3\lambda\sigma^2 & 2g^2\phi \sigma \\ 
2g^2\phi \sigma & m_\phi^2+g^2\sigma^2, \end{array}\right) 
\end{equation}
which for two extreme values of $\sigma$, $\sigma =m_\sigma/\sqrt{\lambda}$
and $\sigma =0$, becomes
\begin{equation}
m^2(\sigma=m_\sigma/\sqrt{\lambda},\phi=0)=\frac{1}{2}\left( \begin{array}{cc}
2m_\sigma^2 & 0\\ 0 & m_\phi^2+g^2m_\sigma^2/\lambda \end{array}\right) 
\end{equation}
and
\begin{equation}
m^2(\sigma=0,\phi)=  \frac{1}{2}   \left( \begin{array}{cc}
-m_\sigma^2 + g^{2}\phi^2 & 0\\ 0 & m_\phi^2  \end{array}\right).
\end{equation}
We see that although the main oscillation frequency scale is $m_\sigma$, since
$\phi$ can be as large as $\phi_{c}\equiv m_\sigma/g$ and since typically
$m_\phi \ll m_\sigma$, the actual frequency scale for the oscillation will be a
weighted time average,whose value can be significantly lower than $m_{\sigma
}$. Let us call this weighted average frequency scale $m_{\sigma }f$, where
$f<1$ is some constant (typically $f$ is of order $10^{-3})$.  As far as the
Hubble expansion rate at the end of inflation is concerned, in the model we
study it is given by
\begin{equation}
H_I \equiv \sqrt{\frac{2 \pi}{3\lambda} \frac{m_\sigma^4}{\mpl^2}}
	= 1.8\times10^{14}\left(\frac{m_\sigma}{10^{-3}\mpl}\right)^2
	        \left(\frac{10^{-2}}{\lambda}\right)^{1/2} \textrm{GeV} .
\label{hend}
\end{equation} 
Then, the ratio
\begin{equation}
\frac{H_I}{fm_{\sigma }}=\sqrt{\frac{2\pi }{3\lambda}}\ 
\frac{m_\sigma}{f\mpl}
\label{eq:manyoscillate}
\end{equation} 
implies that unless $m_\sigma$ is within a factor of $f\sqrt{\lambda}$ of
$\mpl$, many oscillations will occur during the one Hubble time when particle
production occurs.  Hence, the constraint on our parameter space due to
limitations of our background field model is that $m_\sigma$ be as close to as
$\mpl$ as possible.  Since Planckian energy densities invalidate semi-classical
gravitational physics, we will set $m_\sigma$ at the GUT scale,
\begin{equation}
m_\sigma=10^{-3} \mpl,
\label{eq:gutscale}
\end{equation} 
assuming that there is some physics separating the GUT scale and the quantum
gravity scale.  Hence, the following interesting set of parameters ($g=0.01$,
$\lambda =1$, $m_\sigma=10^{-7}\mpl$, $m_{\phi }\approx 652$ GeV) which
satisfy all the constraints and give a mass scales in the intermediate scale
$(10^{12}$ GeV) and the electroweak scale, cannot be analyzed in our model
because in this case, $H_I/(fm_\sigma)$ is too small.  In fact, even for the
single oscillation case, there may be some other damping factor for
$\dot{\phi}$ and $\dot{\sigma}$ which affects the magnitude of $\dot{H}$, which
of course is crucial for the particle production calculation (as we will
explain further in the next section).  Hence, we consider even the numerical
calculation results in this article to be only order of magnitude accurate.

Before we map out the parameter space for which our calculation explicitly is
valid, we would like to show that having $m_\sigma$ close to $\mpl$ forces the
scalar fields to have Planck scale vevs.  This is noteworthy.  Because of the
possible sensitivity to unknown Planck-suppressed operators, scenarios in which
the inflaton attains a Planck scale vev may be unattractive \cite{lyth}.  

We can model the dynamics of $\phi$ before reaching $\phi_c$ as the evolution
of a non-interacting inflaton in a de Sitter background:
\begin{equation}
\ddot{\phi }+3H\dot{\phi }+m_{\phi }^{2}\phi =0.
\end{equation}
For the inflaton field to be slow rolling (overdamped) to the critical value
$\phi _{c}$ from some initial value of $\phi (t=0)>\phi _{c}$, we
must have $m_{\phi }/H\ll 1$. In that case, taking the least
damped solution, we have 
\begin{equation}
\phi =\phi _{c}\exp\left[\frac{1}{3}\left(\frac{m_\phi}{H_I}\right)^{2}
H_I\left(t_{c}-t\right)\right].
\end{equation}
Note that since $\phi_c=m_\sigma/g$, having $m_\sigma$ close to the Planck
scale means that $\phi_c$ will be close to the Planck scale.  We can be more
quantitative by seeing what the constraint $\phi (t=0)< c\mpl$ with $c$ of
order unity implies.  Since $a(t)/a(t=0)=\exp(H_It)$, to have $60$ e-folds, we
must have $\phi(t=0) > 
\phi_{c}\exp\left[60\left(m_{\phi}/H_I)^2/3\right)\right]$.  This implies
\begin{equation}
\frac{m_\phi\mpl}{m_\sigma^2}\sqrt{\frac{30\lambda}{\pi}} < 
\sqrt{\ln \left(\frac{cg\mpl}{m_\sigma}\right)},
\end{equation} 
where we have taken $\phi _{c}=m_\sigma/g$.  There are instances when this
constraint becomes independent of other constraints.  For example,
$g=10^{-4},\, \lambda =1,\, m_{\sigma }=10^{-4}\mpl,\, m_{\phi }=1.7\times
10^{9}$ GeV satisfies all other conditions but this one with $c=1$.  We will
neglect this ``small field'' constraint since this is not as fundamental as
other constraints.

\begin{figure}
\centering \leavevmode\epsfxsize=350pt \epsfbox{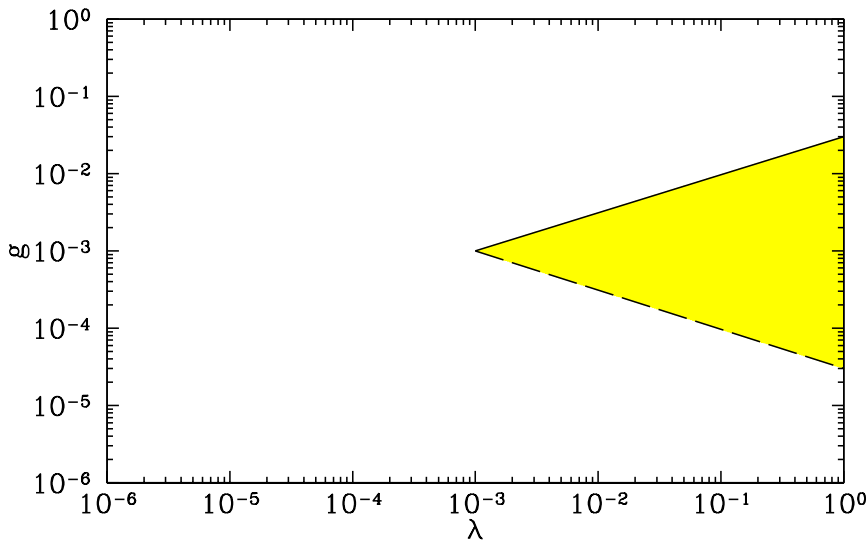}
\caption[]{\label{lambda} The $\lambda$-$g$ parameter space in hybrid 
inflation.  The shaded region corresponds to values of the
parameters allowed by Eqs.\ (\ref{mphi_inequality} -
\ref{other_mphi_inequality}).  The lower limit on this region is dashed
because it represents the "$\gg$" limit in Eq.\ (\ref{g_constraint}).}
\end{figure}

In summary, the parameter space that we will explore will be
\begin{equation}
\label{g_constraint}
\frac{3 \times 10^{-5}}{\sqrt{\lambda}} \ll g \simlt 3 \times
10^{-2} \sqrt{\lambda}
\end{equation}
The parameter space is shown explicitly in Fig.\ \ref{lambda}.

\section{analytic estimate of particle production}

In the appendix we present a general method of estimating particle production
from strong gravitational fields. In this section we apply the results from the
appendix to the hybrid inflationary scenario.  

We show in the appendix that an estimate of particle production requires an
estimate of the background equation solutions.  To start off, let us examine
the time variation of $\dot{H}/H$. After inflation as the scalar fields
oscillate about their minima, $\dot{H}/H$ oscillates. For the envelope of the
function describing the oscillations we have
\begin{equation}
\dot{H}=- \frac{4\pi}{\mpl^2}(\dot{\phi}^2 + \dot{\sigma}^2)\approx 
- \frac{4\pi}{\mpl^2}\rho 
\end{equation}
and the Friedmann equation,
\begin{equation}
H^2=\frac{8\pi}{3\mpl^2}\rho.
\end{equation}
From these, we find a following general relationship after the end of inflation
\begin{equation}
\left. \dot{H} \right|_{\textrm{envelope}}\sim H^2.
\end{equation} 
In fact, after the first oscillation the scalar fields will undergo damped
oscillation about their new minimum, and the scale factor during that time
varies in general as 
\begin{equation} 
a(t)=a_{e}\left(\frac{t}{t_e}\right)^\alpha
\label{aalpha}
\end{equation}
where in the
hybrid inflationary case, $\alpha \approx 2/3$ (which is a typical result of
massive scalar field oscillation).  In reality, this $\alpha$ will have
corrections coming from the phase transition physics. 

Before inflation ends, the scale factor will be taken to evolve as
\begin{equation}
a(t)=a_{e}\exp\left[H_I\left(t-t_e\right)\right]
\label{aoft}
\end{equation}
with $H_I$ given by Eq.\ (\ref{hend}).

Let us now follow the procedure outlined in the appendix to calculate
$n_X(t_e)$. First, consider the contribution to modes that are nonrelativistic
at the end of inflation, $I_a(k)+I_b(k)$, given in Eq.\ (\ref{betaI}).
Assuming $H_I$ is a constant and $a(t)$ evolves as Eq.\ (\ref{aoft}),we find
\begin{equation}
I_a(k)=\frac{1}{4}\ \frac{M_X^2}{k^2} \ a_e^2 \left[ 
	e^{2H_I\left(t_2-t_e\right)} - 
	e^{2H_I\left(t_1-t_e)\right)} \right]  ,
\end{equation}
where, from the Appendix $t_1$ and $t_2$, are defined by
\begin{eqnarray}
k_\textrm{physical}(t_1) = \frac{k}{a_e}\ \frac{a_e}{a(t_1)}=2H_I \nonumber \\
k_\textrm{physical}(t_2) = \frac{k}{a_e}\ \frac{a_e}{a(t_2)}=M_X . 
\end{eqnarray}
Hence, we obtain for $I_a(k)$ the result
\begin{equation}
I_a(k)=\frac{1}{4}\left[1-\left(\frac{M_X}{2H}\right)^2\right] .
\end{equation}

Next, we calculate the nonrelativistic contribution in the period after
inflation, $I_b(k)$, also defined in Eq.\ (\ref{betaI}).  Since $t_{4_i}$ and
$t_{3_{i+1}}$ are close together, ($t_{3_{i+1}}-t_{4_{i}}\ll 1/H$) for
$M_X/H\ll 1$, and since we are concerned with order of magnitudes, we can just
integrate from $t_{3_{1}}$ to $t_{4_{N}}$ instead of summing over each
$i$. Since the nonadiabatic region begins at around $t_e$, we take
$t_{3_{1}}\approx t_e$.  The final integration time, $t_{4_{N}}$, is defined by
the condition $|\dot{H}/H| > M_X$.  In the period after inflation we will take
$a(t)\propto t^\alpha$ as in Eq.\ (\ref{aalpha}), so $H=\alpha/t$ and
$\dot{H}/H=1/t$.  Hence, we have
\begin{equation}
I_b(k) = \frac{1}{2}\int_{t_e}^{t_{4_N}}\ H(t) \ dt = \frac{\alpha}{2}
		\ln\left(\frac{t_{4_N}}{t_e}\right) 
       = \frac{\alpha}{2}\ln \left( \frac{H_I}{M_X\alpha} \right) .
\end{equation}

The calculation of the production of modes relativistic at the end of
inflation, $J_a(k)+J_b(k)$ given in Eq.\ (\ref{betaJ}), is a bit
trickier. First of all, consider the contribution $J_a(k)$:
\begin{equation}
J_a(k) = \frac{1}{2}\int_{t_5}^{t_6} dt \ \frac{a^2(t)}{q^2}H(t) , 
\end{equation} 
where $t_5$ is the time during inflation when $k_\textrm{physical}(t_5)=2H_I$.
During inflation $a(t)=a_e\exp\left[H_I(t-t_e)\right]$, and
$k_\textrm{physical}(t_5) = 2H_I$ gives
\begin{equation}
\frac{k}{a(t_5)}=k a_e\exp\left[-H_I(t_5-t_e)\right] = 2H_I . 
\end{equation} 
The time $t_6$ is the smallest of the times after inflation when
$k_\textrm{physical}(t_6)=2H$ or $k_\textrm{physical}(t_6)=M_X$.  In the period
after inflation, $a(t)=(t/t_e)^\alpha$ and $H=\alpha/t$, so
\begin{equation}
\frac{t_6}{t_e} = \textrm{MIN} \left[ 
\left(\frac{q/a_e}{2H_I/M_X}\right)^{1/(\alpha-1)},
\left(\frac{q}{a_e}\right)^{1/\alpha} \right] ,
\label{eq:choice}
\end{equation}
where the first term is $k_\textrm{physical}(t_6)=2H$ and the second term is
$k_\textrm{physical}(t_6)=M_X$. 

Since $t_6$ will occur after inflation, $J_a(k)$ divides into the parts before
and after inflation:
\begin{eqnarray}
J_a(k) & = & \frac{1}{2}\frac{H_I}{q^2} \int_{t_5}^{t_e} dt \ a^2(t) +
    \frac{1}{2}\frac{1}{q^2} \int_{t_e}^{t_6} dt \ a^2(t)H(t) \nonumber \\
& = & 
\frac{1}{4} \left[ 1 - \left(\frac{M_X}{2H_I}\right)^2 \right] 
\theta \left[\left(\frac{2H_I}{M_X}\right)^\alpha-\frac{q}{a_e}\right]
	\nonumber \\
&  & + \frac{1}{4}\left[ \frac{\left(2H_I/M_X\right)^{2\alpha/(1-\alpha)}}
			      {(q/a_e)^{2/(1-\alpha)}}
 - \left(\frac{M_X}{2H_I}\right)^2\right]
     \theta \left[\frac{q}{a_e}-\left( \frac{2H_I}{M_X}\right)^\alpha\right]
     \theta \left(\frac{2H_I}{M_X}-\frac{q}{a_e}\right).
\end{eqnarray}
where $\theta$ is a step function. The second theta function in the second term
ensures that $t_6/t_e>1$.  Note that $J_a$ matches $I_a$ in the limit
$q/a_e\rightarrow 1$.

To calculate $J_b$, we follow the similar procedure as we did for $I_b$, and
integrate from $t_{7_{1}}=t_6$ to $t_{8_{N}}=t_{4_{N}}$. Note that this is
nonzero only when $t_{7_{1}}\leq t_{8_{N}}$.  Hence, we have
\begin{equation} 
J_b = \left[ \frac{\alpha}{2}\ln \left(\frac{H_I}{M_X\alpha}\right) 
-\frac{1}{2}\ln \left(\frac{q}{a_e}\right) \right] 
\theta \left[\left(\frac{H_I}{\alpha M_X}\right)^\alpha-\frac{q}{a_e}\right]
\theta \left[\left(\frac{2H_I}{M_X}\right)^\alpha-\frac{q}{a_e}\right].
\end{equation}
The first $\theta$ comes from $t_{7_{1}}<t_{8_{N}}$ and the second
$\theta$ comes from using $t_{7_{1}}=t_{6}=t_{e}\left(q/a_e\right)^{1/\alpha}$
(see Eq.~\ref{eq:choice}). Note that $t=t_e\left(q/a_e\right)^{1/\alpha}$
is the time at which the momentum becomes nonrelativistic, and it is precisely
this regime during which $J_b$ is calculated. If $q/a_e > 
\left(2H_I/M_X\right)^\alpha$, then the momentum becomes relativistic and 
there is no extra contribution to $J_b$. From now on we will assume that
$\alpha \geq 1/2$, in which case the second $\theta$ function is irrelevant.

Writing
\begin{equation}
\beta_q=(I_a+I_b)\, \theta(1-q/a_e+\epsilon) + (J_a+J_b)\, \theta(q/a_{e}-1)
\end{equation}
where the $\epsilon$ indicates that we take the first term when $q/a_e=1$, we
can finally obtain the number density $n_X(t_e)$ through
\begin{eqnarray}
n_X(t_e) & = & \frac{M_X^3}{2\pi^2} \int d\left(\frac{q}{a_e}\right) 
\left( \frac{q}{a_e}\right)^2 \left| \beta_q\right|^2 \nonumber \\
& = & \frac{H_I^3z^3}{2\pi^2}\left[ \frac{1}{48}\left(1-\frac{z^2}{4}\right)^2
+\frac{\alpha^2}{12} \ln^2(\alpha z)+A_1+A_2+B_1\right] , 
\end{eqnarray}
where $z\equiv M_X/H_I$ and $\alpha=2/3$ in our case. The first two terms are
the nonrelativistic contribution, and the relativistic contributions $A_i$
and $B_1$ are 
\begin{eqnarray} 
A_{1} & = & 
\frac{(z^2-4)^2}{768} \left[ \left(\frac{2}{z} \right)^{3\alpha}-1 \right]
\nonumber \\   
& = & \frac{0.08}{z^2} - 0.06 - 0.02 z^2 + 0.001z^4 \qquad 
(\textrm{for }\alpha=2/3), \\
A_{2} & = & \frac{1}{768 (9\alpha^2 -1)} \left\{64 z + \left[-24
(1+3\alpha)(1-\alpha)  z^2 + (1+3 \alpha)(1-3\alpha) z^4 \right. \right. 
\nonumber \\ 
& &  \left. \left. -48
(1-3\alpha)(1-\alpha) \right]\left(\frac{2}{z} \right)^{3\alpha} \right\}
\nonumber \\
& = & \frac{0.03}{z^2} - 0.04  
+ 0.03 z - 0.005z^2  \qquad (\textrm{for }\alpha=2/3),
\end{eqnarray}
and
\begin{eqnarray}
B_1 & = & \frac{1}{54}\left[\left(\frac{1}{\alpha z}\right)^{3\alpha}-1\right]
+\frac{\alpha \ln (z)}{18} - \frac{\alpha^2 \ln(\alpha)\ln(z)}{6}
-\frac{\alpha^2\ln^2(z)}{12} +\frac{\alpha\ln(\alpha)}{18}
-\frac{\alpha^2\ln^2(\alpha)}{12} \nonumber \\
& = & \frac{0.04}{z^2} - 0.04 + 0.07 \ln(z)- 0.04\ln^2(z) \qquad 
(\textrm{for }\alpha=2/3) .
\end{eqnarray}

We have neglected cross terms as well since we have neglected any phase
information ({\it i.e.,} if $\beta _q=J_a+J_b$, then $|\beta _q|^2$ was
taken to be $J_a^2+J_b^2$, which should give a lower bound and the
correct order of magnitude since both $J_a$ and $J_b$ are positive). The
important result is that for small $z$, one can approximate
\begin{equation}
\frac{1}{48}\left(1-\frac{z^2}{4}\right)^2 
+ \frac{\alpha^2}{12}\ln^2(\alpha z) + A_1 + A_2 + B_1
\approx \frac{0.15}{z^2}.
\end{equation}
In the $z<1$ limit, the largest contribution comes from the $J_a(k)$ and
$J_b(k)$ terms.  This corresponds to production of modes that are relativistic
at the end of inflation, with approximately equal contributions to the final
value of $|\beta|^2$ coming just before and just after the end of inflation.
We see how the exact behavior of $\dot{H}/H$ after inflation is important.

Finally, putting everything together, in the limit $z=M_X/H_I \ll 1$:
\begin{eqnarray}
\Omega _Xh^2 & \approx & \Omega_Rh^2 \left( \frac{T_{RH}}{T_0}\right) 
\frac{8\pi}{3}\frac{zn_X(t_e)}{\mpl^2H_I}  \nonumber \\
& \approx & \left(\frac{M_X}{10^{11}\textrm{GeV}}\right)^2
            \left(\frac{T_{RH}}{10^9\textrm{GeV}}\right)
             \qquad (\textrm{general result}) \nonumber \\
& \approx &  2 \times 10^4 \left( \frac{T_{RH}}{10^9\textrm{GeV}}\right)
\left( \frac{m_\sigma}{10^{-3}\mpl}\right)^4\frac{z^2}{\lambda}
 \qquad (\textrm{hybrid inflation}) , 
\label{hybridprod}
\end{eqnarray}
where the expression is valid only if $M_X<H_I = 1.8\times10^{14}
(m_\sigma/10^{-3}\mpl)^2(10^{-2}/\lambda)^{1/2}\textrm{GeV}$.  As shown in the
next section, the numerical results corroborate this analytic estimate.

Note that with $M_X\sim10^{13}\textrm{GeV}$ and $T_{RH}\sim10^4\textrm{GeV}$,
we have $\Omega_Xh^2$ of order $10^{-1}$.  Characteristic of gravitational
production, it is possible to produce dark matter many orders of magnitude in
excess of $T_{RH}$.

We have left the $\alpha$ dependence in most of the expressions to indicate
that the mass scaling is sensitive to the fact that the scalar fields enter a
regime just after inflation in which the scale factor evolves as a
matter-dominated universe. The physics of the spinodal decomposition is
expected to change this effective $\alpha$, but one would generically expect
$\alpha$ somewhere between $1/2$ and $2/3$, which means that the number density
of particles produced will roughly remain the same.  Hence, even though all of
our calculations have some sensitivity to more than one oscillations (as can be
seen in our estimation procedure), as long as the scale factor enters a scaling
regime at the end of inflation, our results will give the correct order of
magnitude.

\section{numerical calculation of particle production}

In this section we describe the results of our numerical analysis of
gravitational particle production in the hybrid inflation model.  The basic
hybrid potential was given in Eq.\ (\ref{Linde_potential}).  As discussed
above, the end of inflation is triggered by some perturbation, which we model
by adding to the basic potential a ``perturbed'' potential given in Eq.\
(\ref{eq:perturb}).  The first issue is whether our results are sensitive
either to the nature of the end of inflation or the way we model it.

\begin{figure}
\centering \leavevmode\epsfxsize=350pt \epsfbox{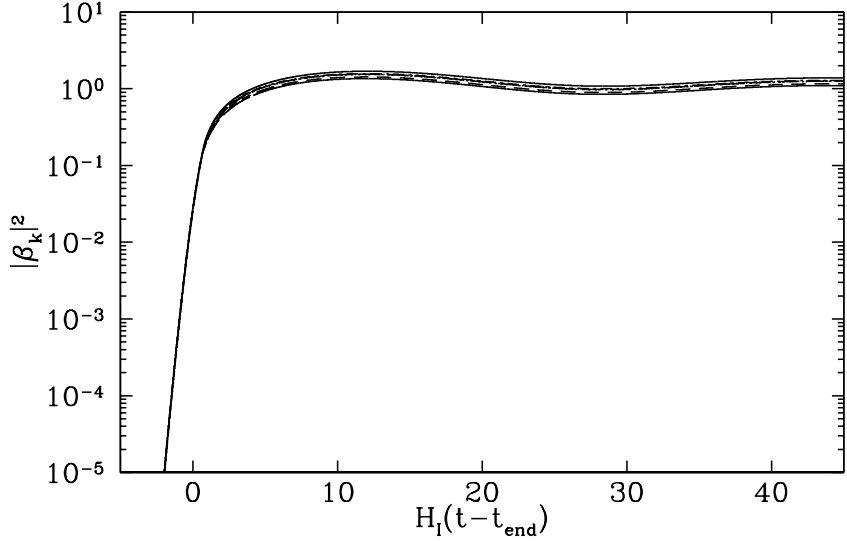}
\caption[]{\label{comp_fig} The absolute square of the Bogoliubov
coefficient as a function of time for several different values of $B$ and $C$
($B$ is dimensionless and $C$ is in units of $\mpl^2$.)  We have $g = 0.01$,
$\lambda = 1$, $M_X = 0.1 H_0$, and $k = 0.1 a_i H_0$.  The lines correspond to
the deformation parameters $(B,C)$ given by $(1,10^9)$; $(10^3,10^7)$;
$(10^5,10^3)$; $(10^5,10^5)$; $(10^5,10^7)$; $(10^5,10^9)$.  The lines are hard
to distinguish on this scale and asymptotically approach within 10\% of each
other.}
\end{figure}

A straightforward exercise is to investigate the sensitivity of particle
production to the parameters $B$ and $C$ in the perturbed potential we use to
trigger the end of inflation.  In Fig.\ \ref{comp_fig} we show the time
evolution of the Bogoliubov coefficient for different choices of $B$ and $C$.
As shown in the figure, our results are insensitive to $B$ and $C$ as long as
they are chosen so as to make $V_P$ negligible outside a very small region
around $\phi = \phi_c$.  Note that we also set $m_\sigma = 10^{-3} \mpl$ for
all the numerical work.

The fact that the final results are insensitive to the exact values of $B$ and
$C$ suggests (but of course does not guarantee) that gravitational particle
production in hybrid inflation will be independent of the mechanism that
triggers the end of inflation.

\begin{figure}
\centering \leavevmode\epsfxsize=350pt \epsfbox{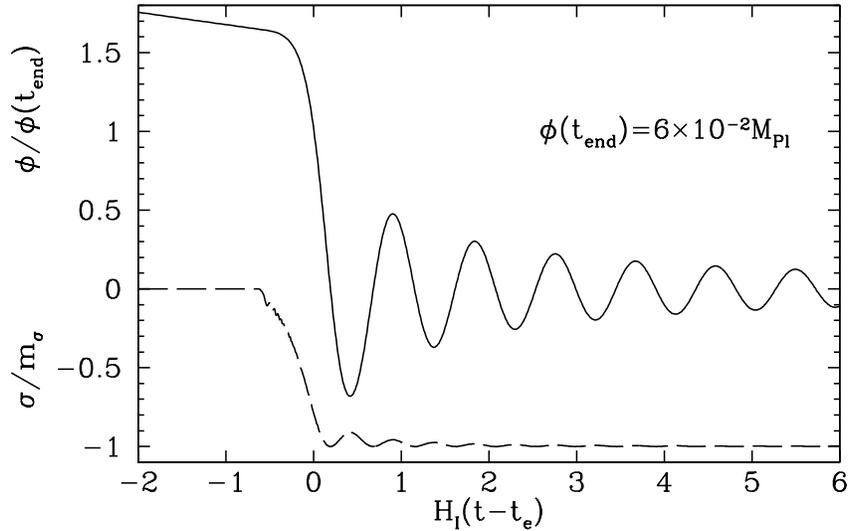}
\caption[]{\label{phi_sigma}  An example of the evolution of the inflaton 
field $\phi$ (solid) and $\sigma$ (dashed) as a function of time at the end of
hybrid inflation. The parameters chosen were $g = 0.01$ and $\lambda = 1$.}
\end{figure}

The evolution of the background fields $\sigma$ and $\phi$ determine the
expansion rate and the change in the expansion rate.  Figure \ref{phi_sigma} is
an example of the evolution of the two fields in hybrid inflation.  For the
parameters of this model ($g=0.01$, $\lambda=1$), the critical value of $\phi$
is $\phi_c=0.1\mpl$.  An instability in the trigger field $\sigma$ (driven by
the ``perturbed'' potential) causes $\sigma$ and $\phi$ to evolve rapidly to
their minima ($\phi=0$, $|\sigma| = m_\sigma/\sqrt{\lambda}=1$) once
$\phi<\phi_c$.

\begin{figure}[p]
\centering \leavevmode\epsfxsize=350pt \epsfbox{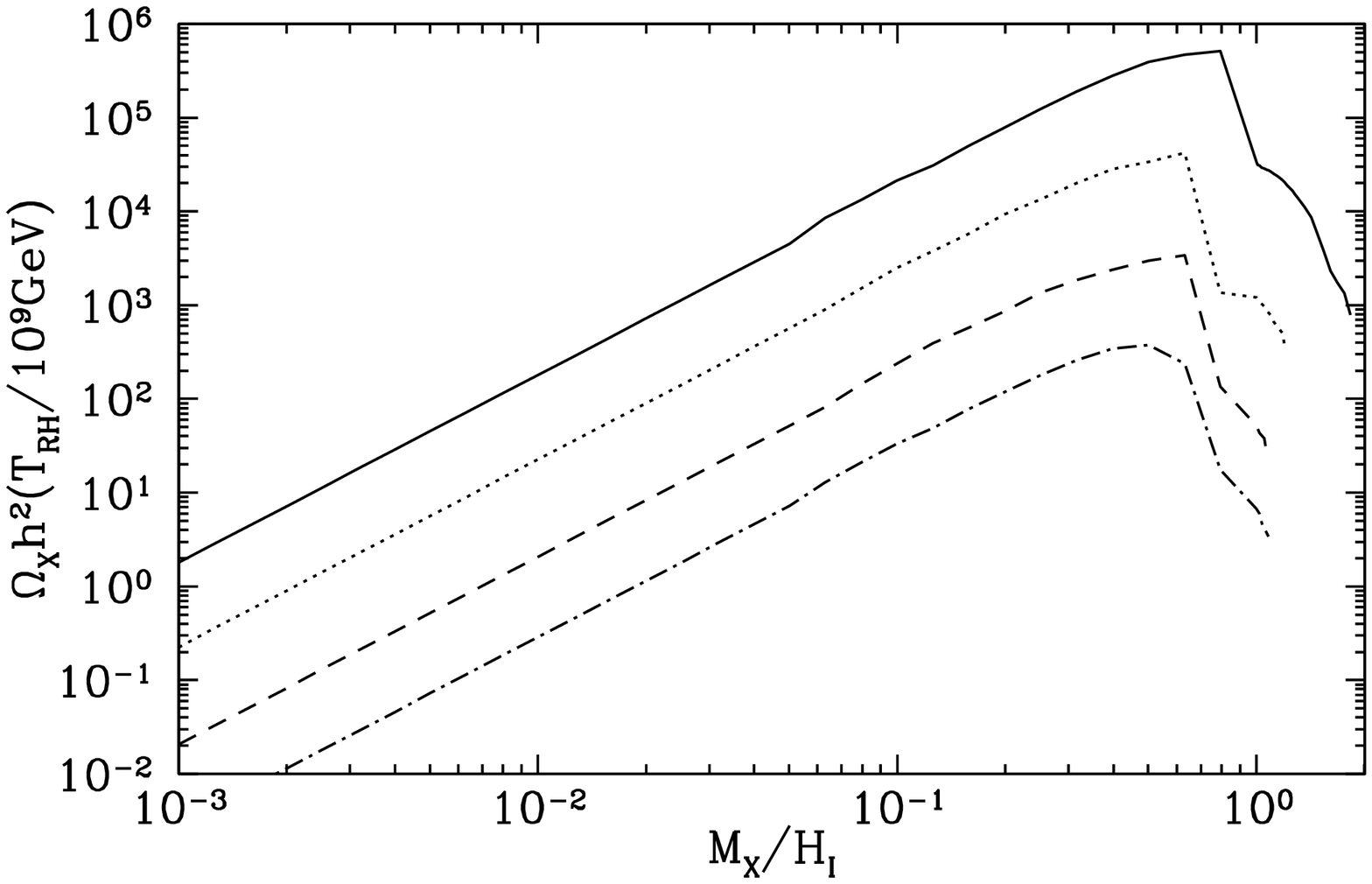}
\caption[]{\label{g0.001_comp} The gravitational production of particles during
hybrid inflation as a function of $\lambda$, with $g$ set to $0.001$.  The
curves correspond to $\lambda$ as follows: solid, $\lambda = 0.001$; dots,
$\lambda = 0.01$; dashes, $\lambda = 0.1$; dash-dot, $\lambda = 1$.  The
magnitude of $\Omega_Xh^2 (T_{RH}/10^9\textrm{GeV})^{-1}$ scales roughly as
$\lambda^{-1}$.}
\centering \leavevmode\epsfxsize=350pt \epsfbox{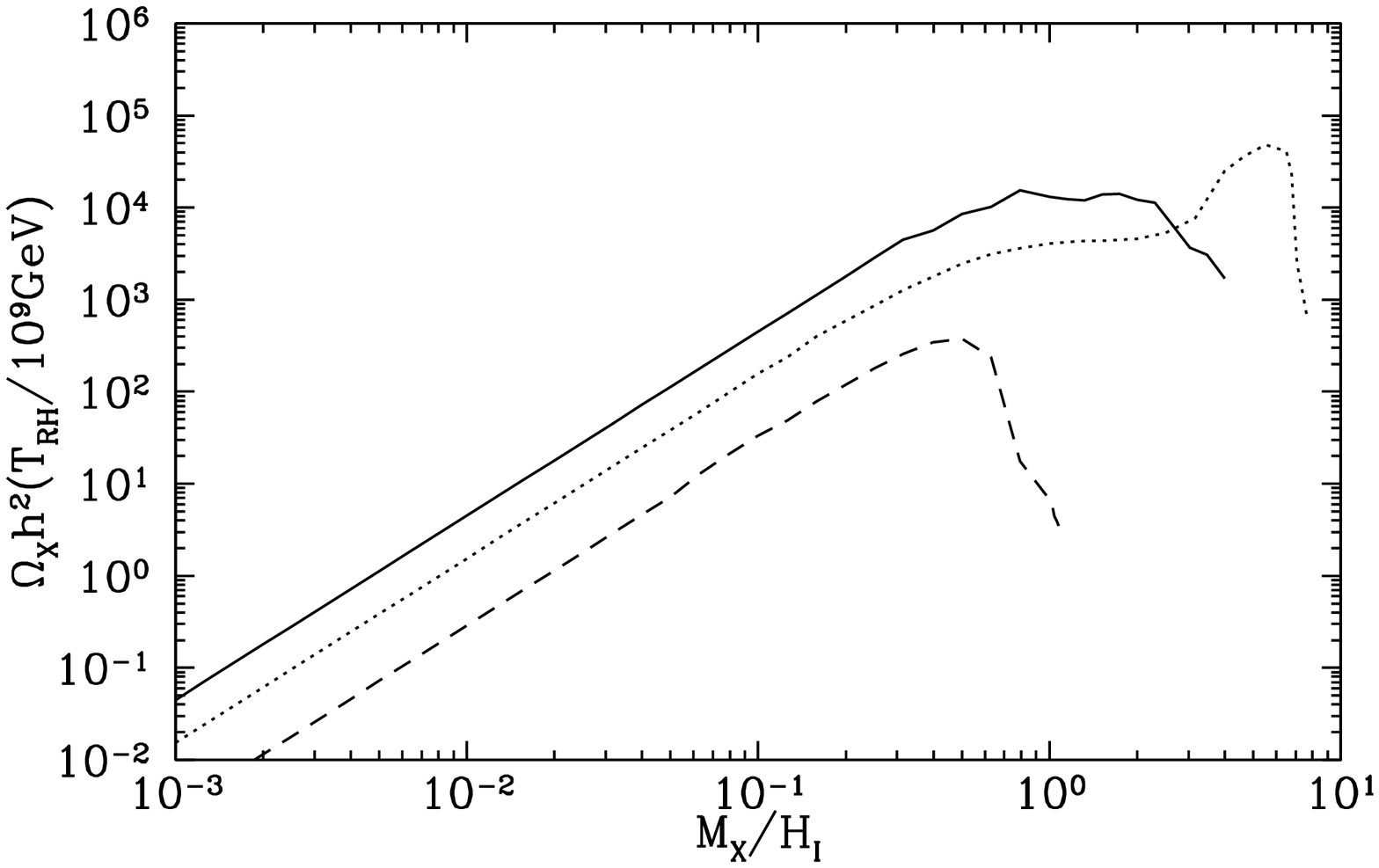}
\caption[]{\label{l1_comp} The gravitational production of particles during
hybrid inflation as a function of $g$, with $\lambda$ set to $1$.
The curves correspond to $g$ as follows: solid, $g = 1$ (note that
this is outside the allowed region of $g,\lambda$ parameter
space); dots, $g = 0.01$; and dashes, $g = 0.001$.}
\end{figure}
\begin{figure}
\centering \leavevmode\epsfxsize=350pt \epsfbox{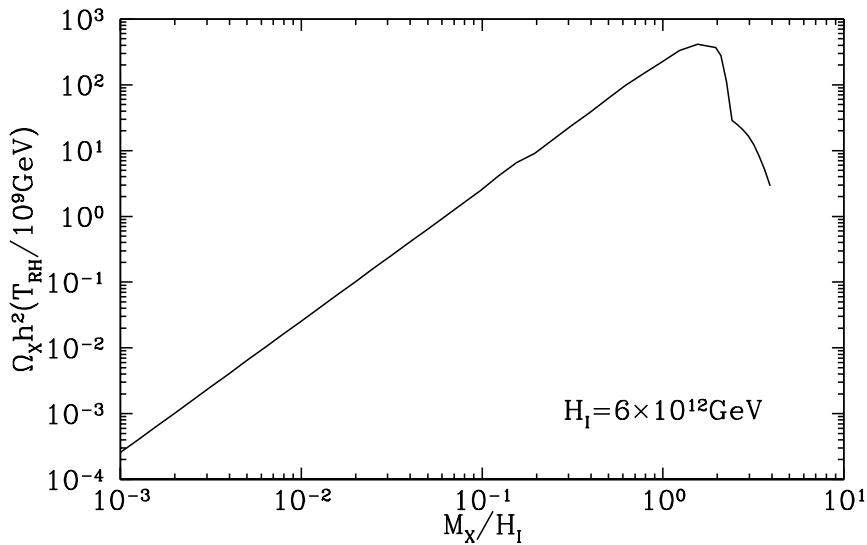}
\caption[]{\label{Natural_spec}  gravitational production of particles
during natural inflation, with $\Lambda = 10^{-3} \mpl$ and
$f_\phi = 0.6 \mpl$.}
\end{figure}

To calculate the relic density of stable particles produced gravitationally,
we integrated the background and X-particle mode equations for several
different points within the allowed regions of parameters shown in Fig.\
\ref{lambda}, as well as for $\lambda = g = 1$, which is well outside it.  Our
results are summarized in Figs.\ \ref{g0.001_comp} and \ref{l1_comp}.

All the curves look similar in form to the mass spectrum for chaotic inflation
with a potential $V(\phi) \sim m_\phi^2 \phi^2$.  The value of $\Omega_Xh^2$
increases with $z=M_X/H_I$ for $z < 1$, then decreases exponentially for
$z>1$.  The reason for this behavior is discussed in this paper for the
small-$z$ region, and in \cite{dcthesis} in the large-$z$ limit.

The numerical results are in qualitative agreement with the result of Eq.\
(\ref{hybridprod}).  

As another example of a single-field model, in Fig.\ \ref{Natural_spec}, we
show the mass spectrum for natural inflation \cite{natural}.  In natural
inflation the potential is usually chosen to be
\begin{equation}
\label{Natural_eq}
V(\phi) = \Lambda^4 \left[1 - \cos\left(\phi/\sqrt{2}f_\phi\right)\right] .
\end{equation}
Normalizing the parameters to produce the observed temperature fluctuations, a
reasonable choice of parameters is $\Lambda=10^{-3}\mpl$ and $f_\phi=0.6\mpl$.
With these choices, $H_I=5.1\times10^{-7}\mpl$.

As in the hybrid inflation case, in the low-$z$ limit $\Omega_Xh^2\propto
M_X^2$.  Again, the numerical results are reasonably represented by
$\Omega_Xh^2\sim (M_X/10^{11}\textrm{GeV})^2(T_{RH}/10^9\textrm{GeV})$.

\section{Conclusions}

The expansion rate of the universe during inflation, $H_I$, may signal a new
mass scale in physics.  The particle spectrum of this new mass scale is
completely unknown.  There may be no particles with this new mass scale; an
example of such a model is $\phi^4$ chaotic inflation.  There may be only one
particle with this mass scale; for example, the inflaton mass in $\phi^2$
chaotic inflation.  Nevertheless, it is very reasonable that one might expect a
rich spectrum of particles of this mass scale.  If this is the case, there may
be nearly stable particles of this mass scale \cite{models}.  Independently of
the coupling of the stable particle, they will be produced as a result of the
expansion of the universe acting on vacuum quantum fluctuations.  It was shown
in Refs.\ \cite{ckr1,ckr2,kuzmin} that such particles would be excellent
candidates for dark matter.  Since the dark-matter particle would have a much
larger mass than usual thermal \WIMPS, they have been named \WIMPZILLAS.

The \WIMPZILLA\ scenario for dark matter seems to be quite robust.  The
\WIMPZILLA\ may be minimally coupled or conformally coupled, it may be a boson
or a fermion, it may couple to the inflaton or may be uncoupled.

The sensitivity of \WIMPZILLA\ production to the inflation model is one of the
subjects of this paper.  Previous calculations have employed a chaotic
inflation model.  Here, we extend our studies on \WIMPZILLA\ production to
hybrid models and natural-inflation models.  We have also developed analytic
techniques that should provide reasonable estimates for \WIMPZILLA\ production
in the limit that $M_X<H_I$.

The general picture for \WIMPZILLA\ production now emerges, and it seems to be
relatively insensitive to the inflation model.  The characteristic expansion
rate during inflation, $H_I$, controls the maximum mass that efficiently can be
produced.  In all inflation models with continuous $\dot{H}$, the production of
particles with mass larger than $H_I$ is exponentially suppressed. For
particles of mass smaller than $H_I$, the contribution to $\Omega_Xh^2$ is
$(M_X/10^{11}\textrm{GeV})^2(T_{RH}/10^9\textrm{GeV})$.

This last expression for $\Omega_Xh^2$ well illustrates that \WIMPZILLA\ masses
much in excess of the reheat temperature may be dark matter.  For instance, if
$T_{RH}=10^4$GeV, then $M_X=10^{13}$GeV would give $\Omega_Xh^2$ in the
desirable range.

While interesting behavior after inflation like preheating or spinodal
decomposition in the case of hybrid inflation might change the results, we
expect the order of magnitude estimate to be correct, and for it to be an
underestimate of \WIMPZILLA\ production.

\begin{acknowledgments}
DJHC was supported by the Department of Energy.  EWK was supported by the
Department of Energy and NASA under Grant NAG5-7092.
\end{acknowledgments}

\appendix \section{analytic determination of particle production}

Consider a minimally coupled scalar field with mass $M_X$.  The equation of
motion for the field is
\begin{equation}
\ddot{X} + 3H \dot{X} - \frac{1}{a^2}\nabla^2 X + M_X^2 X = 0 ,
\end{equation}
where $H$ is the expansion rate.  The scalar field may be expressed in terms of
Fourier modes $X_k=h_k/a$ ($a$ is the scale factor) as
\begin{equation}
X=\int \frac{d^3\!k}{(2\pi)^{3/2}a} \left[ a_k e^{i\vec{k}\cdot\vec{x}}h_k(t)
   + a_k^\dagger e^{-i\vec{k}\cdot\vec{x}}h_k^*(t)\right] ,   
\end{equation}
with the usual normalization condition of the creation and annihilation
operators, $\left[ a_{\vec{k}},a^\dagger_{\vec{l}}\right] = 
\delta^3(\vec{k}-\vec{l})$, the mode functions $h_k$ satisfies the equation 
\begin{equation}
\ddot{h}_k + H \dot{h}_k + \left[ -H^2 - \frac{\ddot{a}}{a} + 
\left(\frac{k}{a}\right)^2 +M_X^2 \right] h_k = 0.
\end{equation}

In terms of Bogoliubov coefficients $\alpha_k$ and $\beta_k$, the mode
functions can be written as  
\begin{equation}
h_k = \frac{\alpha_k}{\sqrt{2\omega_k}} 
      \exp\left(-i\int \omega_k \ a^{-1}(t) \ dt \right)
    + \frac{\beta_k}{\sqrt{2\omega_k}} 
      \exp\left(-i\int \omega_k \ a^{-1}(t) \ dt \right) ,
\end{equation}
where $\omega_k^2=k^2+M_X^2a^2$. Solving for the mode functions is equivalent
to solving the system
\begin{eqnarray}
\dot{\alpha}_k & = & \frac{\dot{\omega_k}}{2\omega_k} 
\exp\left(2i\int \omega_k\  a^{-1}(t) \ dt \right) \beta_k \nonumber \\
\dot{\beta}_k & = & \frac{\dot{\omega_k}}{2\omega_k} 
\exp\left(2i\int \omega_k \ a^{-1}(t) \ dt \right) \alpha_k .
\end{eqnarray}
The gravitational production of particles can be expressed in terms of
the Bogoliubov coefficient $\beta_k$ as
\begin{equation}
n_{X}=\frac{1}{2\pi^2a^3}\int_0^\infty dk\ k^2\left|\beta_k\right|^2
     =\frac{M_X^3}{2\pi^2a^3}\int_0^\infty dq\ q^2\left|\beta_q\right|^2 ,
\end{equation}
where $q \equiv k/M_X = k_\textrm{physical} a/M_X$ with $k$ the comoving
momentum and $k_\textrm{physical} = k/a$ the physical momentum.

The Bogoliubov coefficient to leading adiabatic order can be expressed as (see
Ref.\ \cite{dcthesis} and references therein)
\begin{equation}
\label{bogomaster}
\beta_q(t)\approx \int_{-\infty}^t dt' \ \frac{1}{2}
\left[ \frac{H(t')}{1+q^2/a^2(t')} \right] 
\exp\left(-2iM_X \int^{t'}_{-\infty} dt'' \sqrt{1+q^2 /a^2(t'')} \right) .
\end{equation}
This formula breaks down when $|\beta|$ is of order unity (which may occur in
our scenario), but let us use it to estimate the order of magnitude scales.

The magnitude of $\beta_q$ depends on the magnitude of the argument of the
exponential in Eq.\ (\ref{bogomaster}).  If the argument is of order unity or
greater, then the oscillatory behavior will damp $\left|\beta_q\right|$.  Thus,
the final magnitude of $\beta_q$ depends on the size of $q/a(t)$.  This leads
to a natural division of particle production into the cases where
$q/a(t)=k_\textrm{physical}/M_X$ is larger or smaller than unity.  We will
consider the two cases in turn.

First consider production of nonrelativistic particles: $q/a(t) =
k_\textrm{physical}/M_X < 1$. This case further splits into two subclasses.

$M_{X} < |\dot{H}/H|$: In this case, the oscillations are not important, and
one simply integrates $H(t)$ to the point that it becomes negligible.

$M_{X} > |\dot{H}/H|$: In this case, the oscillations cancel most of the
contribution to the integrand.

Now consider production of relativistic particles: $q/a(t) =
k_\textrm{physical}/M_X > 1$. In this case, the frequency of the oscillations
just becomes the physical momentum.  Again, this case divides into two
subclasses.

$k_\textrm{physical} < |\dot{A}/A|$, where $A(t) \equiv
H(t)/\left[1+q^2/a^2(t)\right]$: Since $q/a(t)>1$, this is equivalent to
$k_\textrm{physical} < \left| 2H + \dot{H}/H\right|$.  In this case the
oscillations are not important, and one simply integrates $H(t)/[q^2/a^2(t)]$
to the point that it becomes negligible.

$k_\textrm{physical} > \left| 2H+\dot{H}/H \right|$: In this case the
oscillations cancel most of the contribution to the integral.

We will neglect the marginal case of $q/a=1$, since this will be roughly
accounted for in the estimates of the above cases.  The different cases and
subcases are given in Table \ref{tablernr}.

\begin{table}
\caption{This table summarizes the different cases in the analytic calculation
of gravitational production of particles. \label{tablernr}}
\begin{tabular}{cccc}
\hline \hline
relativistic/nonrelativistic & subcase & oscillations & $\beta$ \\
\hline
nonrelativistic & $M_X < |\dot{H}/H|$    & none & $\int dt H(t)$ \\
nonrelativistic & $M_X > |\dot{H}/H|$    & many & damped \\
relativistic    & $k_\textrm{physical} < \left|2H+\dot{H}/H\right|$
                            & none & $\int dt H(t)\left[q/a(t)\right]^{-2}$ \\
relativistic    & $k_\textrm{physical} > \left|2H+\dot{H}/H\right|$
                            & many & damped \\
\hline \hline 
\end{tabular}
\end{table}

\begin{figure}
\centering \leavevmode\epsfxsize=350pt \epsfbox{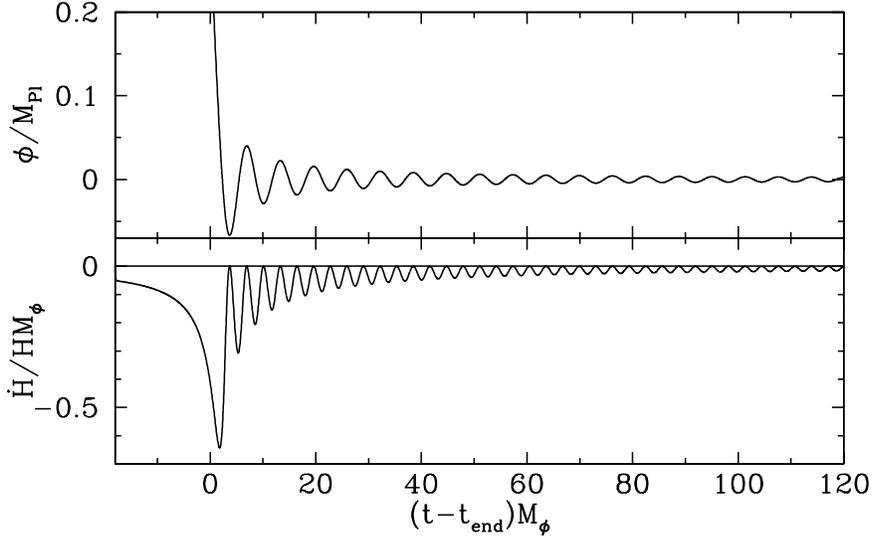}
\caption[]{\label{hdot}  The behavior of $\dot{H}/{H}$ and the inflaton field
at the end of inflation in a simple chaotic inflation model ($V \sim
M_\phi^2\phi^2$).  Notice the oscillatory behavior of $\dot{H}/{H}$ after
inflation.}
\end{figure}

A key to developing analytic approximations is the behavior of $H$ and
$|\dot{H}/H|$.  During inflation, $H$ is roughly constant (denoted as $H_I$)
and $\dot{H}/H \ll H_I$.  After inflation, $\dot{H}/H$ is negative, and
oscillates (with decreasing amplitude) between zero and approximately $-H$.
This behavior is illustrated in Fig.\ \ref{hdot} in the simple chaotic
inflation scenario. During the matter-dominated (MD) phase and the
radiation-dominated (RD) phase, $\dot{H}/H=-3H$ and $\dot{H}/{H}=-4H$,
respectively, so $\left|2H+\dot{H}/H\right|=$[1 (MD) or 2 (RD)]$\times H$

Since $\dot{H}/{H}\simeq 0$ during inflation, from Table \ref{tablernr} we see
that production of nonrelativistic particles is suppressed during inflation and
production of relativistic particles during inflation is suppressed if
$k_\textrm{physical}\gg H$.   

Let us now turn to the estimate of the number density.  The particular
inflation model, together with the behavior of the expansion rate immediately
after inflation, will determine the efficiency of gravitational particle
production.   Here we will give a recipe that can be adapted for several
models. 

We are mainly concerned with the case when $M_X/H_I<1$.  (Particle production
is exponentially suppressed for $M_X/H_I>1$: this case was addressed in detail
in Ref.~\cite{dcthesis}.) The result will depend on whether the particle was
relativistic or nonrelativistic at the end of inflation.

First, consider momentum modes where the particle was nonrelativistic at the
end of inflation, $k_\textrm{physical}(t_e)\leq M_X$. The calculation divides
into production during inflation and post-inflation production.  During
inflation, the growth in $|\beta_k|$ is only when 
$M_X < k_\textrm{physical} < 2H_I$.  After inflation, the particle is
nonrelativistic, and growth occurs during periods when $M_X<\dot{H}/H$.  Using
the results summarized in Table \ref{tablernr} (recall that
$k_\textrm{physical}=k/a=M_Xq/a$),  
\begin{eqnarray}
\beta_k(k_\textrm{physical}(t_e) < M_X) & \simeq & 
	\frac{1}{2}\int_{t_1}^{t_1<t_2\leq t_e} dt \ \frac{H(t)}{q^2/a^2(t)}
   	+ \sum_i\frac{1}{2}\int_{t_{3_i}}^{t_{4_i}} dt \ H(t) \nonumber \\ 
& \equiv & I_a + I_b  .
\label{betaI}
\end{eqnarray}
Here, $I_a$ is the growth during inflation in the interval $\{t_1,t_2\}$ where 
the times are defined such that $k_\textrm{physical}(t_1)=2H_I$ and 
$k_\textrm{physical}(t_2)=M_X$.  $I_b$ is the growth after inflation in the
intervals $\{t_{3_i},t_{4_i}\}$ when $M_X\leq \dot{H}/H$.

Note that we have neglected any phase information between various
contributions.  These interfererence terms should be important in only some
special cases and not generically because in most cases only one term will
dominate.

Now, consider momentum modes where the particle was relativistic at the end of
inflation, $k_\textrm{physical}(t_e)\geq M_X$. Since the mode was relativistic
at the end of inflation, it must have been relativistic throughout inflation.
From Table \ref{tablernr} we see that during inflation, the growth in the
amplitude of $\beta_k$ only occur when $2H_I>k_\textrm{physical}$.  After
inflation, the mode will remain relativistic so long as
$k_\textrm{physical}>M_X$ and it will continue to grow so long as
$2H>k_\textrm{physical}$.  After the mode becomes nonrelativistic
($k_\textrm{physical}<M_X$) it will grow only during periods when
$M_X<\dot{H}/H$. Using the results summarized in Table \ref{tablernr} (recall
that $k_\textrm{physical}=k/a=M_Xq/a$),
\begin{eqnarray}
\beta_k(k_\textrm{physical}(t_e) > M_X) & \simeq & 
	\frac{1}{2}\int_{t_5}^{t_6} dt \ \frac{H(t)}{q^2/a^2(t)}
   	+ \sum_i\frac{1}{2}\int_{t_{7_i}}^{t_{8_i}} dt \ H(t) \nonumber \\ 
& \equiv & J_a + J_b  .
\label{betaJ}
\end{eqnarray}
Here, $J_a$ is the growth during and (possibly) after inflation in the interval
$\{t_5,t_6\}$ where the times are defined such that
$k_\textrm{physical}(t_5) = 2H_I$ and $t_6$ is the smallest of times after
inflation when either $k_\textrm{physical}(t_6)=2H$ or
$k_\textrm{physical}(t_6)=M_X$.  $I_b$ is the growth after inflation
in the intervals $\{t_{7_i},t_{8_i}\}$ when the mode is nonrelativistic and
$M_X \leq \dot{H}/H$.

Once again, we have neglected any phase information between various
contributions for the reason discussed above.

To use these facts to estimate the relic density produce, one must first obtain
a reasonable estimate of $a(t)$ from the background equations.


\end{document}